\documentclass[preprint,aps,pra,showpacs,floatfix]{revtex4-1}
\usepackage[russian]{babel}
\usepackage[utf8]{inputenc}
\usepackage[OT1]{fontenc}
\usepackage{amsmath}
\usepackage{amsfonts}
\usepackage{amssymb}
\usepackage{bm}
\usepackage{graphicx}
\usepackage{booktabs}
\usepackage[left=2cm,right=2cm,top=2cm,bottom=2cm]{geometry}
\usepackage{amsmath}
\usepackage{orcidlink}
\allowdisplaybreaks

\usepackage[labelfont=bf,justification=raggedright,font=small,labelsep=period]{caption}
\usepackage{dcolumn}
\usepackage{siunitx}
\usepackage{multirow}
\usepackage{pgfplots}
\pgfplotsset{compat=newest}
\usepackage{ragged2e}

\usepackage{pgfplots}
\pgfplotsset{compat=newest}

\newcommand{\be}{\begin{eqnarray}}
\newcommand{\ee}{\end{eqnarray}}

\usepackage{color}
\definecolor{BLUE}{rgb}{0.0,0.0,1.0}

\usepackage{soul}
\soulregister\cite7
\soulregister\ref7

\newcommand{\red}{\textcolor{red}}

\begin{document}

\begin{center}

Spectroscopy and Physics of Atoms and Molecules

\vspace*{2mm}

{\large\bf
Axion-Exchange Contribution to the Energy of Lithium-Like Ions
}

\vspace*{2mm}

{\em R.~R.~Abdullin$^{1,2,3,*}$\orcidlink{0009-0004-6992-361X}, A.~V.~Volotka$^{1}$, D.~A.~Glazov$^{1,3}$, M.~G.~Kozlov$^{3,4}$, A.~D.~Moshkin$^{1,2}$, D.~V.~Chubukov$^{2}$} 

$^{1}$School of Physics and Engineering, ITMO University, St. Petersburg 197101, Russia

$^{2}$Saint Petersburg State University, St. Petersburg 199034, Russia

$^{3}$Petersburg Nuclear Physics Institute named after B.P. Konstantinov, National Research Centre “Kurchatov Institute”, Gatchina 188300, Leningrad Region, Russia

$^{4}$Department of Physics, St. Petersburg Electrotechnical University “LETI”, Prof. Popov Str. 5, St. Petersburg 197376, Russia

\end{center}

\textit{*e-mail: rinat.abdullin@itmo.ru}

\vspace*{2mm}

%

\bigskip


\section*{Abstract}
Axions and axion-like particles are among the most promising candidates for dark matter and for manifestations of new physics beyond the Standard Model. In the present work, the contribution of axion exchange to the energy of lithium-like ions is investigated within the framework of relativistic bound-state quantum electrodynamics. A formalism for the interelectronic interaction mediated by axion exchange is developed in the Furry picture with finite nuclear size taken into account. Energy shifts are calculated for a wide range of nuclear charge numbers \(Z\) and axion masses. 
The magnitude of the axion-induced contribution is shown to increase
with increasing \(Z\) for all states considered. Based on the analysis of lithium-like bismuth, constraints on the axion-electron interaction parameters are obtained in the high-mass region. The results indicate that precision spectroscopy of highly charged ions is a promising tool for searches for new physics associated with the exchange of pseudoscalar bosons.

\newpage
\section{Introduction}
In the second half of the 20th century, the Standard Model (SM) was formulated and established, unifying 
the strong, electromagnetic, and weak interactions. 
However, despite the numerous experimental confirmations of its predictions, the SM is not a complete theory: 
it does not include gravity and is 
unable to explain certain 
observed phenomena. For example, it cannot account for cosmological dark energy and dark matter, whose existence is indicated by astrophysical observations. Another unresolved issue is the so-called strong $\mathcal{C}\mathcal{P}$ problem ($\mathcal{C}\mathcal{P}$ denotes the combined charge ($\mathcal{C}$) and parity ($\mathcal{P}$) symmetry): there are no 
experimental manifestations of the violation of this combined symmetry in the quantum chromodynamics (QCD) sector of the SM. At the same time, it is precisely the violation of $\mathcal{C}\mathcal{P}$ symmetry in the early Universe (baryogenesis) that is usually 
used to explain the observed asymmetry between matter and antimatter~\cite{Sakh67}. In this regard, 
an important task of modern theoretical physics is the development of various theories of fundamental interactions beyond the SM (the so-called ``new physics'') and of scenarios for their subsequent verification.

A well-known possible solution to the strong $\mathcal{C}\mathcal{P}$ problem, based on a modification of the QCD Lagrangian, is provided by the Peccei--Quinn mechanism~\cite{Pecc77,Pecc77-2}. 
Shortly after this proposal, Weinberg~\cite{Wein78} and Wilczek~\cite{Wil78} independently showed that the breaking of the global $U(1)$ symmetry proposed in Refs.~\cite{Pecc77,Pecc77-2} should give rise to a pseudoscalar Goldstone boson, later called the axion. 
Subsequent works 
noted that, under certain assumptions about the axion parameters, it can serve as an excellent dark-matter candidate~\cite{Pres83,Abb83,Dine83}. Axion-like particles also arise in various compactifications of string theory~\cite{Svrcek06, Arv10}. One of the key features of axion-like particles appearing in scenarios of new physics is the absence of a strict relation between the particle mass and the strength of its interaction with fermions, in contrast to the QCD axion, for which such a relation is required to
solve the strong $\mathcal{C}\mathcal{P}$ problem. As a result, axion-like particles possess much greater freedom in parameter space, which makes them an important subject of study in various models beyond the Standard Model. Axion-like particles are also promising candidates for cold dark matter, which motivates further investigation of their possible manifestations. For brevity, throughout this paper\red{,} we will refer to axion-like particles simply as axions. Thus, the discovery of axions could shed light on the resolution of the above-mentioned problems of modern fundamental 
interaction physics. 
Numerous astrophysical observations and laboratory experiments have been devoted to the search for axions. Current constraints on axion parameters can be found, for example, in Refs.~\cite{Hare20, bertone05, steffen09, komatsu09, Cong25}.

Atomic physics is an effective tool for testing fundamental theories and searching for new physics~\cite{Saf18, Akul25}. For example, new experiments 
searching for manifestations of axions in atomic and molecular systems have recently been proposed (see Refs.~\cite{Kim23, Agr24} and references therein). Among the types of experiments suitable for imposing 
constraints on axion coupling constants are searches for $\mathcal{T}$-,~$\mathcal{P}$-odd interactions in atomic systems~\cite{Stad18} ($\mathcal{T}$ denotes time-reversal symmetry and is equivalent to combined $\mathcal{C}\mathcal{P}$ symmetry 
due to the well-known $\mathcal{C}\mathcal{P}\mathcal{T}$ theorem). Interpretations of experiments searching for $\mathcal{T}$-,~$\mathcal{P}$-violating effects in terms of axion coupling constants can also be found, for example, in Refs.~\cite{Mais21,Mais21-2,Mais22,Pros23,Pros24}.

In recent decades, unique opportunities have emerged owing to precision experiments with heavy highly charged ions (HCIs). Despite the experimental complexity, HCIs have 
a fundamental advantage over neutral atoms and molecules. In such ions, correlation effects are either absent, as in the one-electron case, or can be calculated with high accuracy. This places new demands on atomic theory, which must 
reliably predict the sensitivity of atomic systems to new interactions.
In the present work, we consider the sensitivity of HCIs to the electron--axion interaction. In contrast to previous studies, we investigate the $\mathcal{C}\mathcal{P}$-even contribution of pseudoscalar axion exchange to the energy of lithium-like ions. This provides a unique opportunity to establish stringent constraints on the axion parameters.

Throughout this paper, relativistic units \(\hbar = c = 1\) are used (\(\hbar\) is the Planck constant). The units of charge are 
\(\alpha = e^2/(4 \pi)\), where \(e<0\) is the electron charge and \(\alpha\) is the fine-structure constant.

\newpage
\section{Relativistic Formalism of Virtual Axion Exchange}

 The interaction Lagrangian of the pseudoscalar axion $(A)$ with the Standard Model electron $(\psi)$ can be written in the following form:
\begin{equation}
\mathcal{L}_{\text{int}} = g_{A} i A \bar{\psi} \gamma_5 \psi.
\label{eq:EQ1}
\end{equation}
In Eq.~(\ref{eq:EQ1}), \(g_A\) is the dimensionless axion interaction constant, and \(\bar{\psi}\) is the Dirac conjugate of the electron-positron field \(\psi\), defined as \(\bar{\psi} = \psi^\dagger \gamma_0\). The axion field \(A\) and the Dirac field \(\psi\) are represented as
\begin{equation}
A(x) = \int \frac{d^3 k}{(2\pi)^3 \sqrt{2\omega}} \left( a_k e^{-ikx} + a_k^\dagger e^{ikx} \right),
\label{eq:EQ2}
\end{equation}
\begin{equation}
\psi(x)=\sum_{\varepsilon_p>0} b_p \psi_p(x)+\sum_{\varepsilon_p<0} c_p^\dagger \psi_p(x).
\label{eq:EQ3}
\end{equation}
where \(a_k^\dagger (c_p^\dagger)\) and \(a_k(b_p)\) are the creation and annihilation operators for axions (positrons and electrons), respectively, \(\psi_p\) are the solutions of the Dirac equation, and \(\varepsilon_p\) is the one-particle energy of the state \(\psi_p\) corresponding to the set of quantum numbers \(p\). Here \(x\) and \(k\) denote 4-vectors, where \(x = (t, \bm{x})\) is the coordinate in Minkowski space, and \(k = (\omega, \bm{k})\) is the wave 4-vector, with \(\omega = \sqrt{\bm{k}^2 + m_A^2}\), where \(m_A\) is the axion mass. The Dirac matrices $\gamma =(\gamma^0,\gamma^1,\gamma^2,\gamma^3)= (\gamma_0,\bm{\gamma})$ are given in the representation in which $\gamma_5 =i\gamma^0\gamma^1\gamma^2\gamma^3 = \tiny{\begin{pmatrix} 0 & 1 \\ 1 & 0 \end{pmatrix}}$, $\gamma_0 = \tiny{\begin{pmatrix} 1 & 0 \\ 0 & -1 \end{pmatrix}}$. In what follows, the above notation is used without additional explanation.

In the present work, the standard method of bound-state quantum electrodynamics (QED) is employed, using 
the wave functions of electrons in an external field, namely, in the Coulomb field of the nucleus, 
as the initial basis. This means that the interaction of electrons with the nuclear field is included already in the zeroth-order approximation, which corresponds to the so-called Furry picture. Finite nuclear size is also taken into account in the calculations.

The energy corrections arising from virtual axion exchange for an arbitrary state of the electronic system \( | B \rangle \) are calculated according to the following formula~\cite{mohr85}:
\begin{eqnarray}
 \Delta E &=& {g_{A}^2} \iint d\bm{r}_1 \, d\bm{r}_2 \sum_{\substack{p_1,\, p_2 \\ p_3,\, p_4}} 
\psi_{p_1}^\dagger(\bm{r}_1)i \gamma_0 \gamma_5 \psi_{p_2}(\bm{r}_1) 
\psi_{p_3}^\dagger(\bm{r}_2)i \gamma_0 \gamma_5 \psi_{p_4}(\bm{r}_2) D(\Delta,\bm{r}_1; \bm{r}_2)  \nonumber \\
&\times & \langle B |b_{p_1}^\dagger b_{p_2} b_{p_3}^\dagger b_{p_4} | B \rangle.
\label{eq:EQ4}
\end{eqnarray}
The matrices \(\gamma_{5}\) correspond to different vertices of the diagram. From the law of 
conservation of energy, it follows that \( \Delta = |\varepsilon_{p_1} - \varepsilon_{p_2}| = |\varepsilon_{p_3} - \varepsilon_{p_4}| \). The axion propagator is defined as follows:
\begin{eqnarray}
D(t,\bm{r}_1; \bm{r}_2) = \int \frac{d^4 k}{(2\pi)^4} \frac{i}{k^2 - {m_{A}^2}+i\epsilon} e^{-i k_0 t + i \bm{k} \cdot \bm{r}}.
\label{eq:EQ5}
\end{eqnarray}
To calculate the contributions to the energy of bound states, we need the propagator in the energy representation. For this purpose, we perform the Fourier transform:
\begin{equation}
D(\omega, \bm{r_1}; \bm{r_2}) =\int dt \int \frac{d^4 k}{(2\pi)^4} \frac{e^{-i k_0 t + i \bm{k} \cdot \bm{r}} e^{i \omega t}}{k_0^2 - \bm{k}^2 - {m_{A}^2}+i\epsilon} = -\frac{1}{4\pi } \frac{e^{ir\sqrt{\omega^2 - m_{A}^2}}}{r}.
\label{eq:EQ7}
\end{equation}
Let us introduce the interelectronic-interaction operator induced by virtual axion exchange: 
\begin{equation}
I_{A}(\omega) = {g_{A}^2}i\gamma_0 \gamma_5 i \gamma_0 \gamma_5D(\omega, \bm{r}_1; \bm{r}_2).
\label{eq:EQ8_1}
\end{equation}
Using the introduced operator, Eq.~(\ref{eq:EQ4}) takes the following compact form:
\begin{eqnarray}
 \Delta E &=& \sum_{\substack{p_1,\, p_2 \\ p_3,\, p_4}} \langle p_1 p_3 | I_{A}(\omega) | p_2 p_4 \rangle \langle B |b_{p_1}^\dagger b_{p_2} b_{p_3}^\dagger b_{p_4} | B \rangle.
\label{eq:EQ8_2}
\end{eqnarray}

Next, let us consider the matrix element of axion exchange between two electrons in the general form:
\begin{equation}
\langle ab | I_{A}(\omega) | cd \rangle = {g_{A}^2}\iint d\bm{r}_1 d\bm{r}_2 \psi^\dagger_a(\bm{r}_1)i \gamma_0 \gamma_5 \psi_c(\bm{r}_1) \psi^\dagger_b(\bm{r}_2)i \gamma_0 \gamma_5 \psi_d(\bm{r}_2) D(\omega, \bm{r}_1; \bm{r}_2).
\label{eq:EQ8_3}
\end{equation}
In the present work, we consider the motion of an electron in the Coulomb field of the nucleus \(V({r})\). This field possesses spherical symmetry. In such a field, there exist electron states with definite values of energy, angular momentum, and parity. The wave functions of such states have the form:
\begin{equation}
\psi_p = \psi_{n\kappa M} = \begin{pmatrix}
g_{\kappa n}(r) \, \Omega_{\kappa M}(\bm{n}) \\
if_{\kappa n}(r) \, \Omega_{-\kappa M}(\bm{n})
\end{pmatrix}.
\label{eq:EQ9}
\end{equation}
Here \(g_{\kappa n}(r)\) and \( f_{\kappa n}(r) \) are the radial functions of the relativistic electron in the Coulomb field of the nucleus, while \( \Omega_{\kappa M}(\bm{n})\) and \( \Omega_{-\kappa M}(\bm{n})\) are the spherical spinors that determine the angular dependence. These functions are solutions of the Dirac equation with the potential \(V({r})\):
\begin{equation}
  \bigl[\boldsymbol{\alpha}\!\cdot\!\mathbf{p} + \beta\, m + V({r})\bigr]\,\psi_{n\kappa M}(\mathbf{r}) = \varepsilon_{n\kappa M}\,\psi_{n\kappa M}(\mathbf{r}),
  \label{eq:EQ10}
\end{equation}
where \(\bm{\alpha} = \gamma_0 \bm{\gamma}\), and \(\beta = \gamma_0\). The eigenvalue \( \varepsilon_{n\kappa M} \) corresponds to the energy of the one-particle state.   
The expansion into spherical harmonics for the explicit form of the propagator reads as follows:
\begin{equation}
\frac{e^{i \eta r}}{r} = 4 \pi i \eta \sum_{J,\, M} j_J(\eta r_<) h_J^{(1)}(\eta r_>) Y_{JM}(\bm{n}_1) Y_{JM}^*(\bm{n}_2),
\label{eq:EQ11}
\end{equation}
where \(j_J(\eta r_<)\) and \(h_J^{(1)}(\eta r_>)\) are the spherical Bessel and Hankel functions of the first kind, respectively, and \(\eta=\sqrt{\omega^2 - m_{A}^2}\).
We should note that, in the case of virtual axion exchange, 
the upper and lower components of the Dirac bispinor are mixed:
\begin{equation}
\psi_a^\dagger \gamma_0 \gamma_5 \psi_c = (i f_a \Omega_{\bar{a}}^\dagger g_c \Omega_c) + (i g_a \Omega_a^\dagger f_c \Omega_{\bar{c}}),
\label{eq:EQ12}
\end{equation}
and a similar mixing is also observed in the Breit contribution associated with photon exchange. Substituting into Eq.~(\ref{eq:EQ11}) the expressions for the axion propagator and the electron wave functions, we obtain 
the matrix element of the operator 
represented as a product of two factors. One of them does not depend on the angular-momentum projections, while the other is expressed in terms of 3j symbols (or Clebsch--Gordan coefficients):
\begin{align}
\langle ab | I_{A}(\Delta) | cd \rangle
&= \frac{g_{A}^2}{4 \pi} \sum_{J,\,M} (-1)^{j_a - M_a + j_b - M_b + J - M}
\nonumber \\
&\quad \times
\begin{pmatrix}
j_a & J & j_c \\
-M_a & M & M_c
\end{pmatrix}
\begin{pmatrix}
j_b & J & j_d \\
-M_b & -M & M_d
\end{pmatrix}
\langle ab \| I_{A}(\Delta) \| cd \rangle_J ,
\label{eq:EQ13}
\end{align}
where the reduced matrix element of the axion-interaction operator is
\begin{eqnarray}
\langle ab || I_{A}(\omega) || cd \rangle_J &=& \iint dr_1 \, dr_2 \, (-1)^J g_J(\omega, r_<, r_>) 
\nonumber \\
 &\times&
[iF_a(r_1) G_c(r_1) C_J(-\kappa_a, \kappa_c) + iG_a(r_1) F_c(r_1) C_J(\kappa_a, -\kappa_c)]   \nonumber \\
 &\times& [iF_b(r_2) G_d(r_2) C_J(-\kappa_b, \kappa_d) + iG_b(r_2) F_d(r_2) C_J(\kappa_b, -\kappa_d)].
\label{eq:EQ14}
\end{eqnarray}
For convenience, we introduced the following notation 
in Eq.~(\ref{eq:EQ14}): 
\begin{equation}
C_J (\kappa_a, \kappa_c) = (-1)^{j_c + \frac{1}{2}} \sqrt{(2j_a+1)(2j_c+1)(2l_a+1)(2l_c+1)}
\begin{pmatrix}
l_a & J & l_c \\
0 & 0 & 0
\end{pmatrix}
\begin{Bmatrix}
j_a & J & j_c \\
l_c & \frac{1}{2} & l_a
\end{Bmatrix},
\label{eq:EQ15}
\end{equation}
\begin{equation}
g_J (\omega, r_<, r_>) = i\eta (2J+1) j_J (\eta r_<) h_J^{(1)} (\eta r_>), \quad F = r f(r), \quad G = r g(r).
\label{eq:EQ16}
\end{equation}
\section{Calculation for Lithium-Like Ions}
Let us consider axion exchange in a lithium-like ion, which is a three-electron system. In accordance with Eq.~(\ref{eq:EQ8_2}), the action of the creation operators on the vacuum state is given by the 
expression: \( | B \rangle \) = \(b_1^\dagger b_2^\dagger b_3^\dagger | 0\rangle \), and thus, the energy corrections for a three-electron system are:
\begin{eqnarray}
 \Delta E &=& \sum_{\substack{p_1,\, p_2 \\ p_3,\, p_4}} \langle p_1 p_3 | I_{A}(\omega) | p_2 p_4 \rangle 
 \langle 0 | b_1 b_2 b_3 b_{p_1}^\dagger b_{p_2} b_{p_3}^\dagger b_{p_4} b_1^\dagger b_2^\dagger b_3^\dagger | 0 \rangle.
\label{eq:EQ17}
\end{eqnarray}
A lithium-like ion contains three electrons, and we need to account for 
the interaction of each electron with the other electrons. 
The energy correction associated with axion exchange between two electrons is determined by the contributions of two different diagrams: the direct and exchange diagrams. Thus, for three electrons, there are six diagrams. 
This can be obtained 
from Eq.~(\ref{eq:EQ17}) by expanding the expressions for the creation and annihilation operators 
using the anticommutation relations. The resulting expression splits into six terms, each corresponding to a diagram of virtual axion exchange between a pair of electrons in the ion. 
A similar situation is observed for the diagrams of virtual photon exchange in a lithium-like ion.

\begin{figure}[!htb]
	\centering
\includegraphics[width=0.5\textwidth]{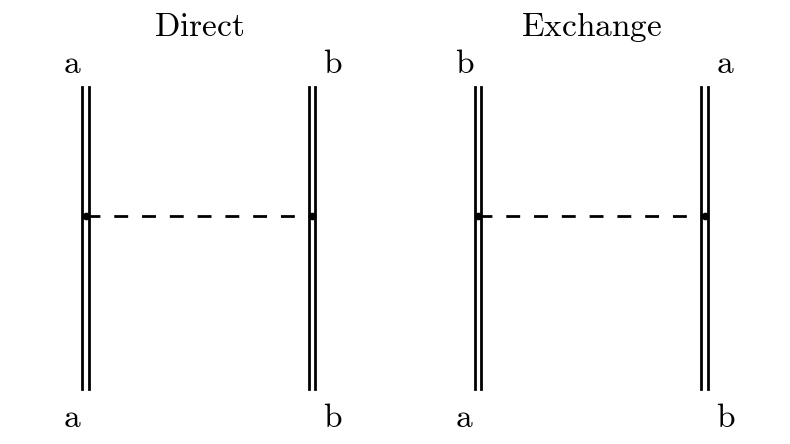} 
	\caption{Feynman diagrams (direct and exchange ones) for axion exchange. The dashed line corresponds to the axion, and the double line to the bound electron.}\label{fig:photo1}
\end{figure}

In a three-electron system, 
the contribution of virtual axion exchange between each pair of electrons should be considered. Let us denote the selected pair of electrons by \(a\) and \(b\). Then, the energy correction corresponding to this pair includes two terms --- the direct and exchange contributions, shown in Fig.~\ref{fig:photo1}. The expression 
for the direct contribution is:
\begin{eqnarray}
\Delta E_{\text{dir}} &=&  \langle ab | I_{A}(0) | ab \rangle 
= \frac{g_{A}^2}{4 \pi} \sum_{J,\,M} (-1)^{j_a - M_a + j_b - M_b + J - M} \nonumber \\
&\times& \begin{pmatrix}
j_a & J & j_a \\
-M_a & M & M_a
\end{pmatrix}
\begin{pmatrix}
j_b & J & j_b \\
-M_b & -M & M_b
\end{pmatrix}  \langle ab || I_{A}(0) || ab \rangle_J.
\label{eq:EQ18}
\end{eqnarray}
where the frequency of the virtual axion is equal to zero.
The exchange contribution has the form:
\begin{eqnarray}
\Delta E_{\text{exc}} &=& \langle ab | I_{A}(\Delta_{ab}) | ba \rangle 
= \frac{g_{A}^2}{4 \pi} \sum_{J,\,M} (-1)^{j_a - M_a + j_b - M_b + J - M} \nonumber \\
&\times& \begin{pmatrix}
j_a & J & j_b \\
-M_a & M & M_b
\end{pmatrix}
\begin{pmatrix}
j_b & J & j_a \\
-M_b & -M & M_a
\end{pmatrix}  \langle ab || I_{A}(\Delta_{ab}) || ba \rangle_J,
\label{eq:EQ19}
\end{eqnarray}
where \( \Delta_{ab} = |\varepsilon_a - \varepsilon_b|\).

\begin{figure}[!htb]
    \centering
    \includegraphics[width=0.75\textwidth]{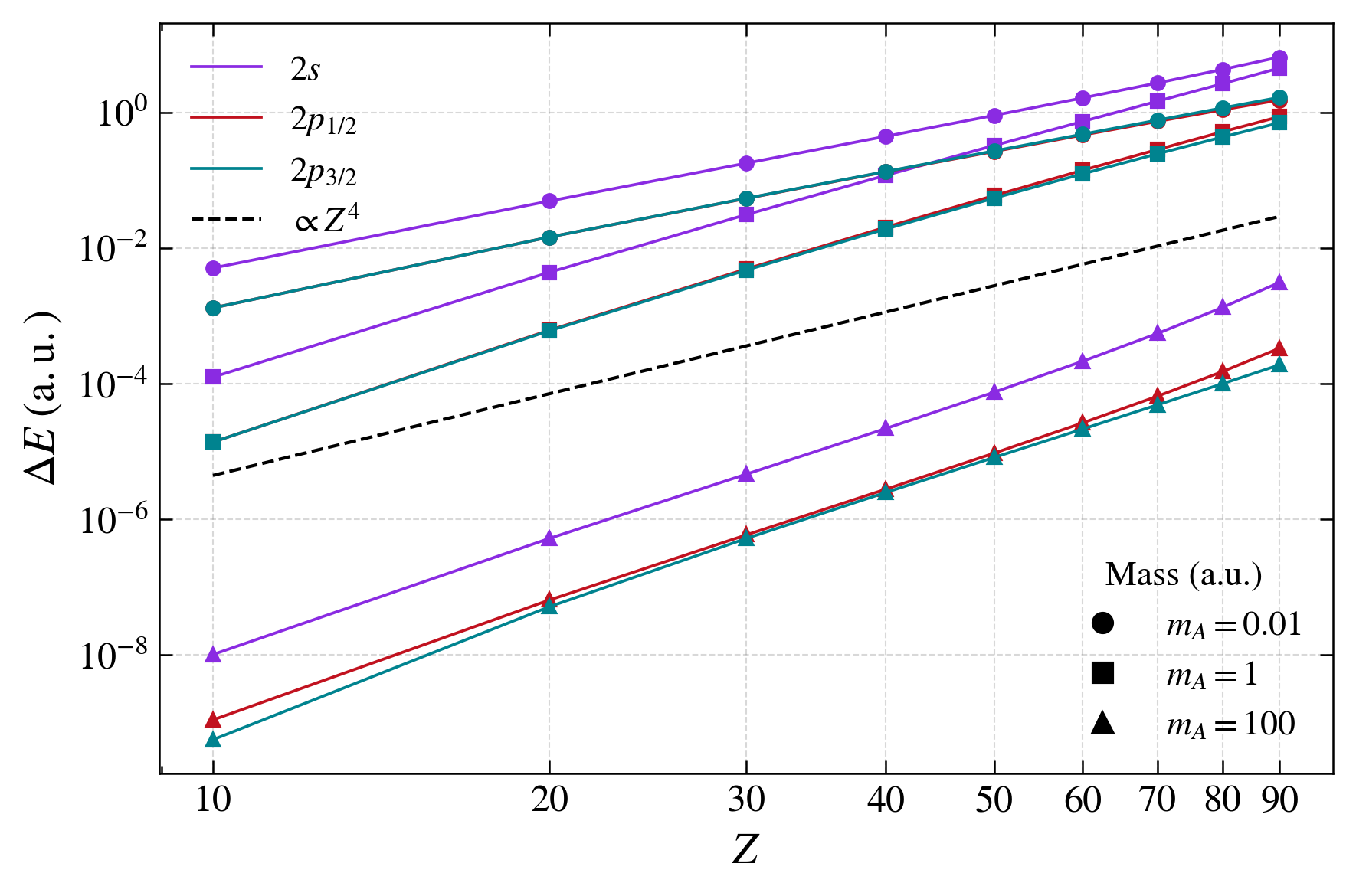}
    \caption{\justifying Ionization-energy correction induced by virtual axion exchange in lithium-like ions as a function of the nuclear charge number \(Z\). Results are shown for the \(\left(1s\right)^2 2s\), \(\left(1s\right)^2 2p_{1/2}\), and \(\left(1s\right)^2 2p_{3/2}\) states for several fixed values of the axion mass. In all calculations, \(\frac{g_A^2}{4\pi}=\alpha\) was assumed.}
    \label{fig:photo2}
\end{figure}

Let us consider specific states, namely\red{,} \(\left(1s\right)^2 2s\), \(\left(1s\right)^2 2p_{1/2}\) 
and \(\left(1s\right)^2 2p_{3/2}\). For these configurations, the ionization-energy corrections induced by virtual axion exchange were calculated numerically as functions of the nuclear charge number \(Z\). The results for several 
values of the axion mass are presented in Fig.~\ref{fig:photo2}. 
We can see that the axion-exchange contribution increases with increasing \(Z\) for all the considered states. 
For comparison, 
the power-law dependence \(\propto Z^4\) is indicated with the dashed line. We should note 
that, in these calculations, the axion interaction constant was chosen to be equal in magnitude to the electron charge, 
i.e. \(\frac{g_A^2}{4\pi}=\alpha\).

 Next, we will analyze 
 the axion-exchange contribution to the ionization energy as a function of 
 the axion mass -- 
 by varying the axion mass and keeping \(Z\) values fixed. 

\begin{figure}[!htb]
    \centering
    \includegraphics[width=0.75\textwidth]{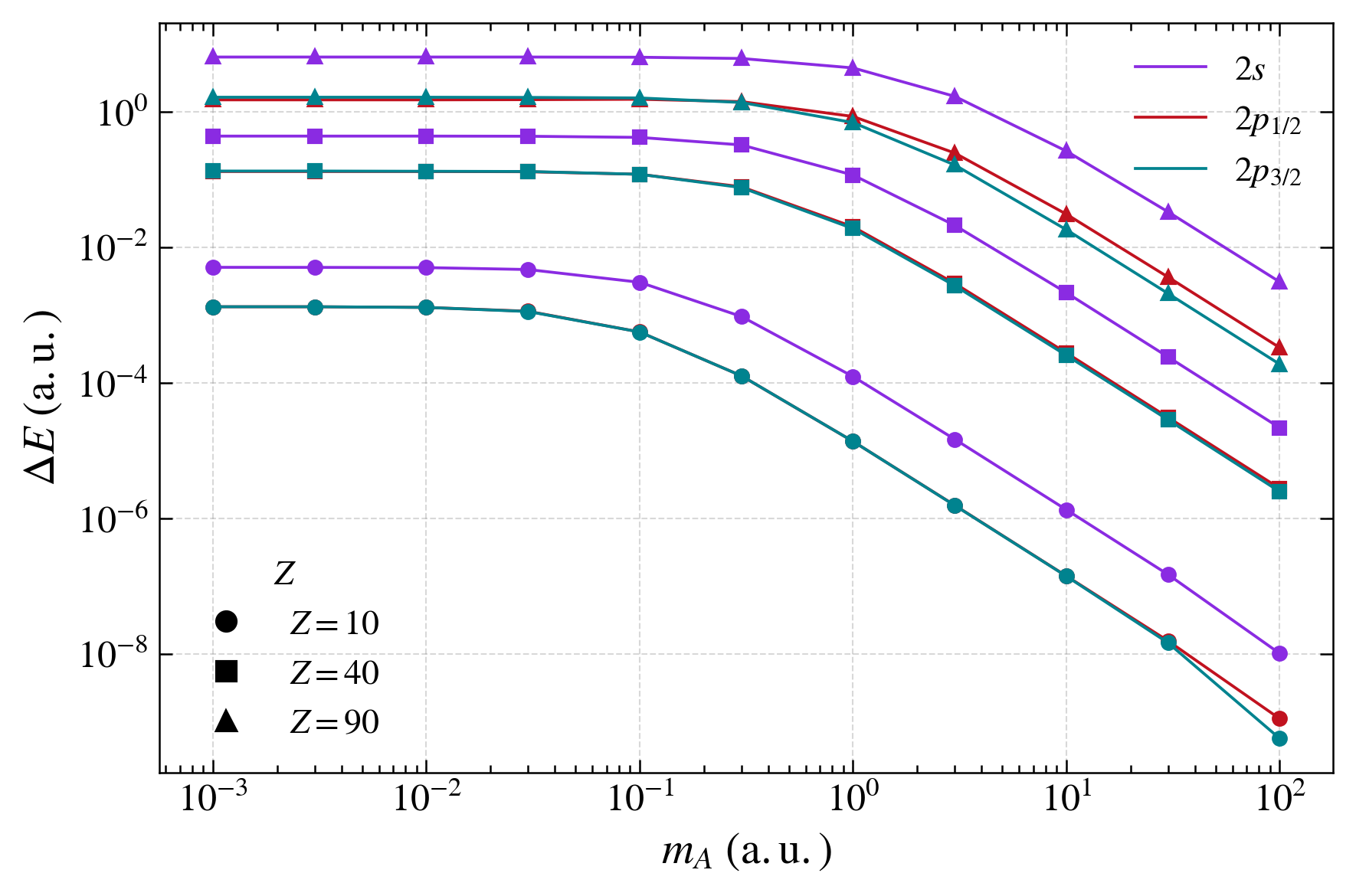}
    \caption{\justifying Dependence of the ionization-energy correction induced by virtual axion exchange in lithium-like ions on the axion mass \(m_{A}\). Results are shown for the \(\left(1s\right)^2 2s\), \(\left(1s\right)^2 2p_{1/2}\), and \(\left(1s\right)^2 2p_{3/2}\) states for several fixed values of the nuclear charge number \(Z\). In all calculations, \(\frac{g_A^2}{4\pi}=\alpha\) was assumed.}
    \label{fig:photo3}
\end{figure}
The dependence of the axion contribution on the axion mass is determined by the characteristic scale of the induced interaction. 
The Compton wavelength corresponding to the axion mass defines 
the effective range of this interaction. As the axion mass increases, the Compton wavelength decreases, and the interaction becomes 
shorter-ranged. For this reason, in atomic systems with larger 
electron shells, the contribution of heavy mediators is suppressed. In contrast, in highly charged ions, the characteristic interelectronic distances are smaller, and these systems remain sensitive 
to heavier axions. As 
Fig.~\ref{fig:photo3} shows, a decrease in the characteristic size of the electron shell 
expands the mass region in which the axion contribution can manifest itself.

Given this 
enhanced sensitivity of highly charged ions to interactions mediated by heavier axions, it is natural to 
consider the constraints on the axion interaction parameters next. For this purpose, the \(\left(1s\right)^2 2p_{3/2} - \left(1s\right)^2 2s\) transition in the lithium-like bismuth ion \(^{209}\mathrm{Bi}\) was chosen. For this transition, in Ref.~\cite{Yerokhin25}, 
the transition energy was calculated with a high precision, 
and the theoretical value was compared with experimental data. The 
theory and experiment agreed sufficiently well 
to use this transition to derive constraints on the axion interaction parameters.
\begin{figure}[!htb]
    \centering
    \includegraphics[width=0.75\textwidth]{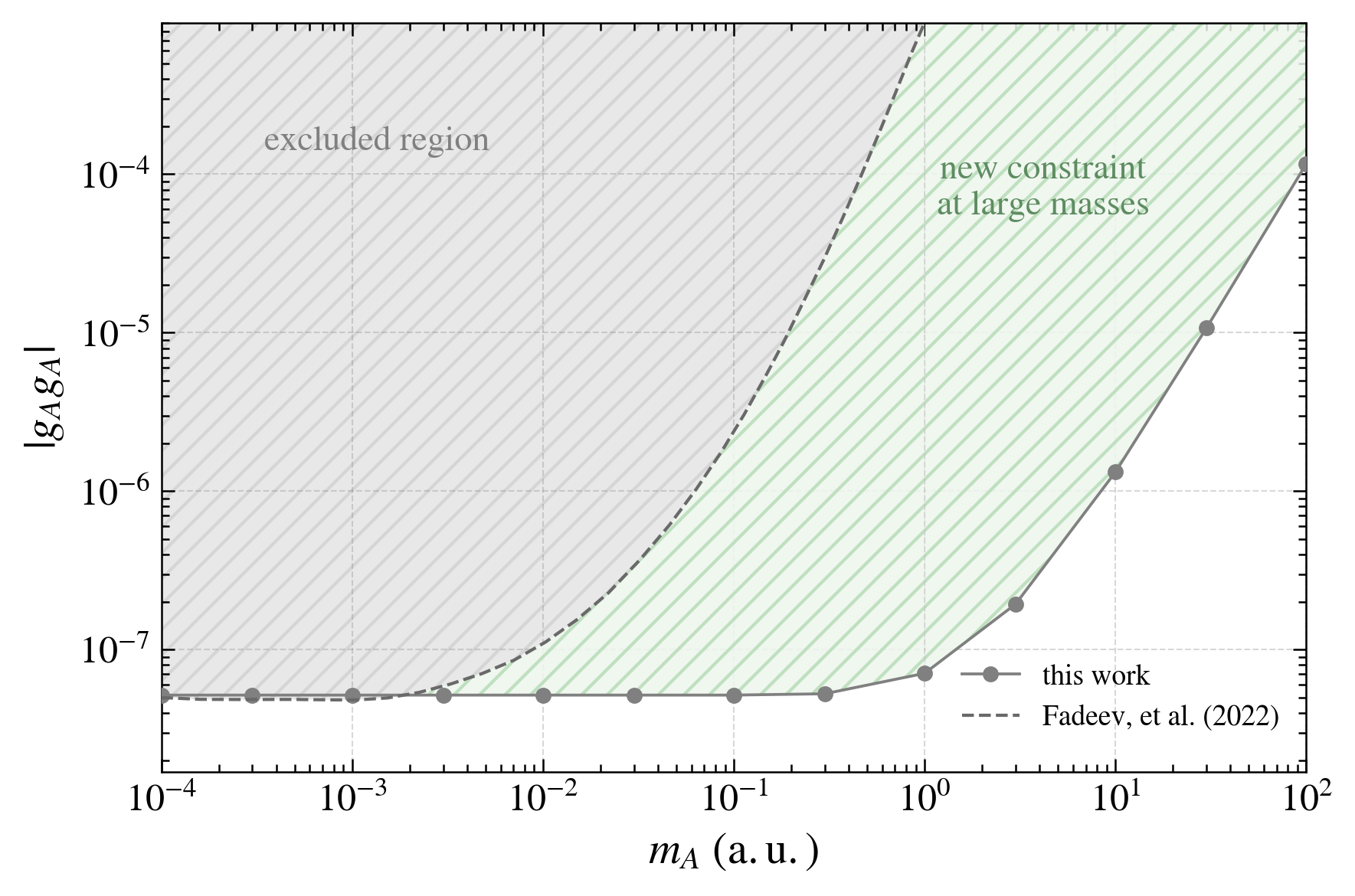}
    \caption{\justifying Constraints on the axion interaction parameters obtained from the comparison of the theoretical and experimental values of the \(\left(1s\right)^2 2p_{3/2} - \left(1s\right)^2 2s\) transition energy in lithium-like bismuth. For comparison, the constraint from Ref.~\cite{Fadeev2022} is also shown.}
    \label{fig:photo4}
\end{figure}

The constraints were determined from the requirement that the additional contribution due to axion exchange should not violate the agreement between the theoretical and experimental values of the transition energy within \(2\sigma\). Accordingly, for each fixed value of the axion mass \(m_A\), the maximum allowed value of the coupling constant was determined at which 
the axion-exchange contribution does not spoil the agreement between theory and experiment within the interval defined by the combined theoretical and experimental uncertainty.

The results presented in Fig.~\ref{fig:photo4} show that the \(\left(1s\right)^2 2p_{3/2} - \left(1s\right)^2 2s\) transition in the lithium-like bismuth ion can be used to obtain a new constraint in the high-mass\red{-}axion region. Compared with the result of Ref.~\cite{Fadeev2022}, the obtained new constraint 
extends into the region of heavier axions while maintaining comparable sensitivity to the axion interaction constant. This confirms that highly charged ions are promising systems for the search for effects induced by the exchange of heavy bosons.

\clearpage 
\clearpage

\clearpage
\begingroup
\setlength{\parskip}{0pt}

\section{Conclusion}
\indent In the present work, the contribution of pseudoscalar axion exchange to the energy of lithium-like ions has been investigated within the framework of relativistic bound-state quantum electrodynamics. A formalism for the interelectronic interaction mediated by axion exchange has been developed in the Furry picture with finite nuclear size taken into account, and expressions for the direct and exchange contributions to the energy of the three-electron system have been obtained.

Based on the developed formalism, we have numerically calculated 
the ionization-energy corrections for the \(\left(1s\right)^2 2s\), \(\left(1s\right)^2 2p_{1/2}\), and \(\left(1s\right)^2 2p_{3/2}\) states 
of lithium-like ions in a wide range of nuclear charge numbers \(Z\) and axion masses \(m_A\). 
The axion-exchange contribution has been shown to increase 
with increasing \(Z\) for all the considered states.

The analysis of the axion-exchange contribution to the ionization energy as a function of 
the axion mass has 
revealed that a decrease in the characteristic size of the electron shell in highly charged ions 
expands the mass range in which the axion contribution remains significant. 
Thus, we establish that heavy lithium-like ions are of particular interest for the search for effects induced by the exchange of high-mass axion-like particles, since the smaller characteristic size of their electron shell makes these ions more sensitive to a shorter-range interaction 
compared to 
ordinary atomic systems.

By comparing 
the theoretical and experimental values of the \(\left(1s\right)^2 2p_{3/2} - \left(1s\right)^2 2s\) transition energy in the lithium-like bismuth ion, constraints on the axion-electron interaction parameters have been obtained. 
We have shown that this transition 
allows reaching the high-mass axion region unavailable 
with previously obtained constraints, 
at the same time retaining competitive sensitivity to the coupling constant.

The obtained results demonstrate that precision spectroscopy of highly charged lithium-like ions is a promising tool for the search for new physics associated with the exchange of pseudoscalar bosons. 
This approach may be further developed 
both with the analysis of other high-precision spectroscopic transitions in highly charged ions and with the search for manifestations of new physics in the bound-electron \(g\)-factor. In particular, 
similar analysis may be applied to the helium atom, for which the 
discrepancy between theoretical and experimental values of the ionization energy is currently under discussion~\cite{Cong26}, while the developed relativistic formalism may serve as a natural foundation 
for the analysis of such a system.
\endgroup
\section*{Funding}

This work was supported by the Russian Science Foundation under Project No.~24-72-10060.
\section*{Conflict of Interest}

The authors declare that they have no conflict of interest.


\renewcommand{\refname}{References}

\end{document}